\pdfoutput=1
\documentclass[aip,prb,twocolumn,graphicx]{revtex4}

\usepackage{graphicx}
\usepackage{caption}

\begin{document}

\title{Magnetic monopole field exposed by electrons} 

\author{Armand B\'{e}ch\'{e}}
\author{Ruben Van Boxem}
\author{Gustaaf Van Tendeloo}
\author{Jo Verbeeck}
\affiliation{EMAT, University of Antwerp, Groenenborgerlaan 171, 2020 Antwerp, Belgium}

\begin{abstract}
The experimental search for magnetic monopole particles \cite{Dirac,Milton,Wu} has, so far, been
in vain. Nevertheless, these elusive particles of magnetic charge have fueled a rich field of
theoretical study \cite{Bonnardeau,Frisch,Aad,Castelnovo,Salomaa,Cardoso,Goddard}. Here, we
created an approximation of a magnetic monopole in free space at the end of a long, nanoscopically
thin magnetic needle \cite{Kasama}. We experimentally demonstrate that the interaction of this
approximate magnetic monopole field with a beam of electrons produces an electron vortex state, as
theoretically predicted for a true magnetic monopole
\cite{Wu,Kasama,Aharonov,Kruit,Bliokh,Uchida,Verbeeck,Fukuhara,Tonomura}. This fundamental quantum
mechanical scattering experiment is independent on the speed of the electrons and has consequences
for all situations where electrons meet such monopole magnetic fields as for example in solids.
The setup not only shows an attractive way to produce electron vortex states but also provides a
unique insight into monopole fields and shows that electron vortices might well occur in
unexplored solid-state physics situations.
\end{abstract}

\maketitle

Magnetic monopoles have provided a rich field of study, leading to a wide area of research in
particle physics \cite{Bonnardeau,Frisch,Aad}, solid state physics \cite{Castelnovo}, ultra-cold
gases \cite{Salomaa}, superconductors \cite{Cardoso}, cosmology \cite{Bonnardeau}, and gauge
theory \cite{Goddard}. As electric charges can be seen as monopole sources and sinks of electric
field lines, the strong symmetry with magnetic and electrical fields e.g. in the free space
Maxwell equations \cite{Berry,Bliokh_dual,Cameron} hints to the possible existence of magnetic
monopoles as well. So far, the search for such magnetic monopoles has been unsuccessful. However,
an effective monopole field can be produced at the tip of a nanoscopic magnetized ferromagnetic
needle \cite{Kasama,Fukuhara}. The Aharanov-Bohm effect \cite{Aharonov} can be used to understand
the effects of such a monopole field on its surroundings which is crucial to its observation and
provides a better grasp of fundamental physical theory. Previous studies have been limited to
theoretical semiclassical optical calculations of the motion of electrons in such a monopole field
\cite{Kruit}. Solid state systems like the recently studied `spin ice' provide a constrained
system to study similar fields, but make it impossible to separate the monopole from the material
\cite{Castelnovo}. Here, we realize the diffraction of fast electrons on the magnetic monopole
field generated by the extremity of a long magnetic needle. Free space propagation of the
electrons helps to understand the dynamics of the electron-monopole system without the complexity
of a solid state system and will allow various areas of physics to use the effects of monopole
fields. Various predictions about angular momentum, paths of travel and general field effects can
readily be studied using the available equipment. The experiment performed here shows that even
without a true magnetic monopole particle, the theoretical background on monopoles serves as a
basis for experiments.

Indeed it has been predicted that when a plane electron wave interacts with a hypothetical
magnetic monopole, a vortex electron state would arise
\cite{Wu,Kasama,Aharonov,Kruit,Bliokh,Uchida,Verbeeck,Fukuhara,Tonomura}:
\begin{equation}
\Psi_{out} = \Psi_{in}\exp(im\phi),
\end{equation}
with $m$ depending on the charge of the magnetic monopole and $\phi$ the azimuthal angle in the
plane perpendicular to the electron wave propagation.

Approximating the magnetic monopole now by the end of a magnetic needle leads to similar effects.
Indeed, such a semi-infinite cylinder of magnetic flux has been considered in earlier work on
magnetic monopoles but has remained a Gedanken experiment so far \cite{Lipkin}. From the
description of the monopole field by a vector potential, a flux line, or `Dirac string' arises as
a mathematical pathology which should be undetectable if a magnetic monopole was to be a true
monopole, leading to the famous magnetic charge quantization \cite{Dirac}.

The magnetic vector potential is a mathematical tool used in quantum physics which has real,
measurable effects \cite{Aharonov} which were experimentally demonstrated by electron diffraction
\cite{Tonomura,Chambers}. The Aharonov-Bohm (AB) phase is acquired by an electron when its path
encloses magnetic flux:
\begin{equation}
\Delta\phi_{AB} = \frac{e}{\hbar c}\oint\mathbf{A}.d\mathbf{s}.
\end{equation}
This phase is a purely quantum mechanical effect as it is present even if the electron does not
cross a region containing magnetic flux, a case where classical forces have no influence on the
passing electrons. The Aharonov-Bohm effect is most often discussed with infinite cylinders of
magnetic flux, avoiding the interesting end points where the magnetic field $\mathbf{B} =
\mathrm{rot} \mathbf{A}$, takes the form of a monopole:
\begin{equation}
\mathbf{B} = \frac{\mathbf{r}}{r^{3}}
\end{equation}
If one calculates the Aharonov-Bohm phase for electrons passing perpendicular by a semi-infinite
cylinder of flux, one obtains a linear azimuthal dependency around the ending point of the
cylinder \cite{Wilczek}:
\begin{equation}
\Delta\phi_{AB} = \frac{2e}{\hbar c}g\phi.
\end{equation}
This means that a passing electron will indeed be transformed into a vortex state:
\begin{equation}
\Psi_{out} = \Psi_{in}\exp(im\phi).
\end{equation}
For a true monopole field, where the charge $g$ is quantized, this leads to an integer m ($g = m
\hbar c/(2e)$), resulting in a perfect phase vortex of topological charge $m$. There are several
different derivations of this phase factor, all extensively discussed in literature, and all
predicting the same vortex phase factor \cite{Wu,Lipkin,Wilczek}. A sketch and discussion on the
subtle differences between the effect of a semi-infinite cylinder of flux and a true monopole is
given in Supp. Fig. 8.

Carefully tuning a magnetic needle leads to the same phase structure which is indistinguishable
from a true monopole as long as the needle is thin and the flux converges towards a quantized
flux. In this letter we successfully produced an approximation to a Dirac string with a nanoscopic
magnetized ferromagnetic needle. The interaction of a plane electron wave with only one end of the
needle allows the typical azimuthal AB phase shift to occur and vortex electron states to be
created, as sketched in Fig. 1a.

\begin{figure}[h]
\begin{center}
\includegraphics[width=\columnwidth]{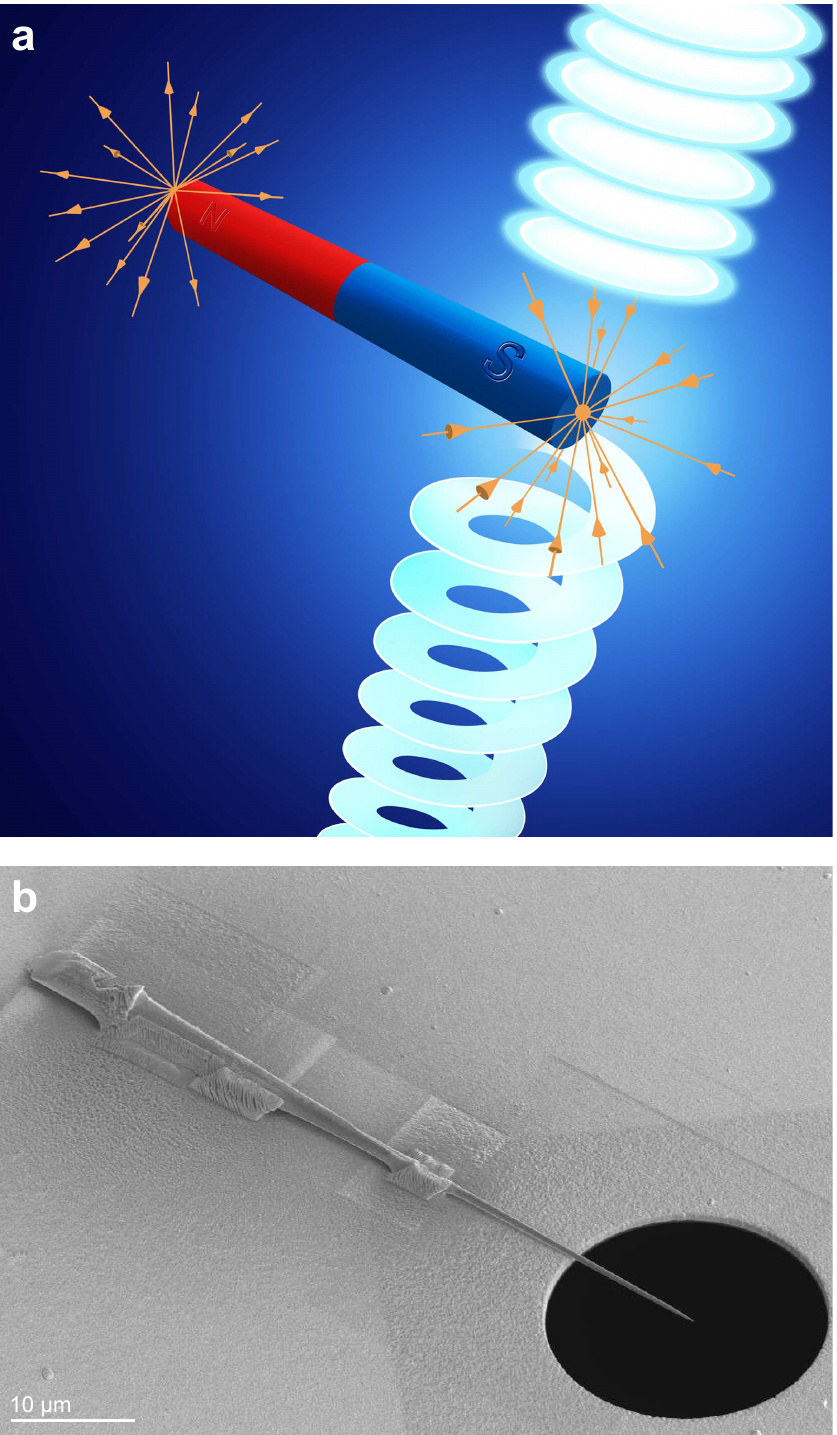}
\caption{\protect\textit{Concept and design of the monopole field: (a) An incoming electron plane
wave is transformed into a vortex beam with a helical wavefront, through interaction with the
magnetic monopole field. (b) SEM view of the experimental design. The nickel needle and its copper
base are soldered to a gold plated SiN aperture using FIB assisted Pt deposition. Half the Ni
needle is positioned over a 20 $\mu$m circular aperture, forming a local monopole
field.}}\label{Figure_1}
\end{center}
\end{figure}

From an experimental point of view, the needle is extracted from bulk Ni making use of a focused
ion beam (FIB) instrument, resulting in a cone approximating an elongated cylinder with a cone
angle of about 2 degrees shown in Fig. 1b. The strong shape anisotropy between the needle length
(21.4 $\mu$m) and the tip diameter of only 200 nm leads to a situation where only a single on-axis
magnetic domain occurs. After shaping the needle, it is positioned over a 20 $\mu$m circular
aperture drilled in a nonmagnetic Au coated thin SiN film, in order to make sure electrons can
only interact with one end of the needle and its magnetic monopole field.

We can verify the magnetic state at the tip of the nickel needle (red square in Fig. 2a) by
inserting it in a transmission electron microscope (TEM) and performing electron holography in
field free conditions \cite{Kasama,Tonomura}, sketched in Supp. Fig. 1. This method measures the
Aharanov-Bohm phase shift of the electrons caused by the magnetic vector potential around the
needle. The resulting experimental phase map is shown in Fig. 2b and reveals the typical spiraling
character in qualitative agreement with a finite element simulation for the same shape given in
Fig. 2(c) and Supp. Fig. 2. The phase image resembles that of optical spiral phase plates, as used
to create optical vortices \cite{Beijersbergen}. Exposing the needle to an external on-axis
magnetic field flips the axis of magnetization without going through multi domain states (Supp.
Fig. 3). When the magnetization direction is reversed, the handedness of the phase reverses as
expected (Supp. Fig. 3). In this sense, the needle tip behaves as a magnetic monopole with a
polarity that can be chosen depending on the magnetization direction.

\begin{figure*}[t]
\begin{center}
\includegraphics[width=2\columnwidth]{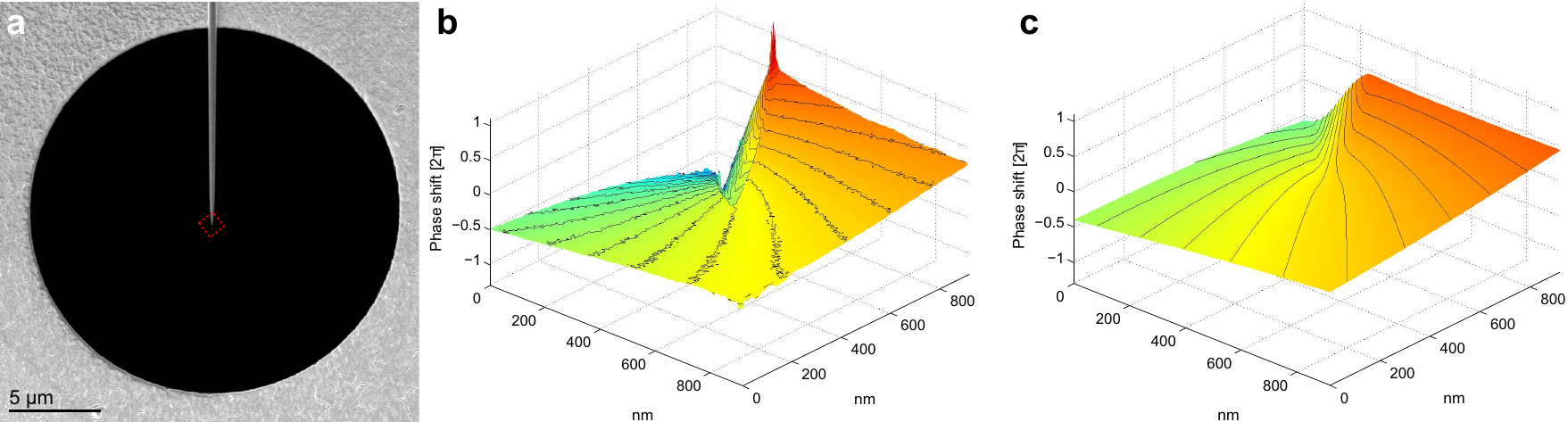}
\caption{\protect\textit{Effect of the needle on the phase of the electrons: (a) Magnified SEM
image of the needle positioned onto the circular aperture. The red dashed square region indicates
the position of images b and c. (b) Experimental phase map caused by the magnetic field around the
Ni needle obtained by electron holography in field free conditions. The phase map is drawn in 3D
to emphasize its helicity. (c) Finite element simulation of the phase map around a model for the
needle. Detailed phase profiles are supplied in Supp. Fig. 4.}}\label{Figure_2}
\end{center}
\end{figure*}

Illuminating the needle with a plane electron wave (300 kV, $\lambda=$1.97 pm) inside a TEM allows
to experimentally verify whether a magnetic monopole field creates a vortex electron state. A
series of images is recorded in the far field at different defocus of an imaging lens showing, in
Fig. 3a, the presence of a central dark region. This persistent area of destructive interference
is a clear sign of a phase discontinuity in the center, as expected for vortex waves. The ring is
not exactly closed which occurs when a non-integer orbital angular momentum is present
\cite{Berry_OpticalVortices}. Decomposing the phase map over the full aperture for the simulated
magnetized needle into OAM eigenmodes indeed indicates that the deviation from a pure cylindrical
shape leads to a distribution of OAM eigenmodes with an average of -5.8$\hbar$ per electron (Supp.
Fig. 5). These experimental observations agree remarkably well with wave optical simulations
presented in Fig. 3b ruling out the possibility that the dark region is caused by a shadowing
effect (see Supp. Fig. 6 for further simulations).

We can also prove experimentally that this electron wave now possesses net orbital angular
momentum, induced by the interaction with the monopole field, making use of the Gouy phase method
\cite{Bliokh_ElectronVortex,Guzzinati}, sketched in Supp. Fig. 1. For waves with net OAM we expect
a $\pi$ rotation of the image when going through focus with a direction of rotation depending on
the sign of the OAM. This exact behavior is observed in Fig. 3c, which shows a clear clockwise
rotation when going from under to over-focus.

\begin{figure*}[t]
\begin{center}
\includegraphics[width=2\columnwidth]{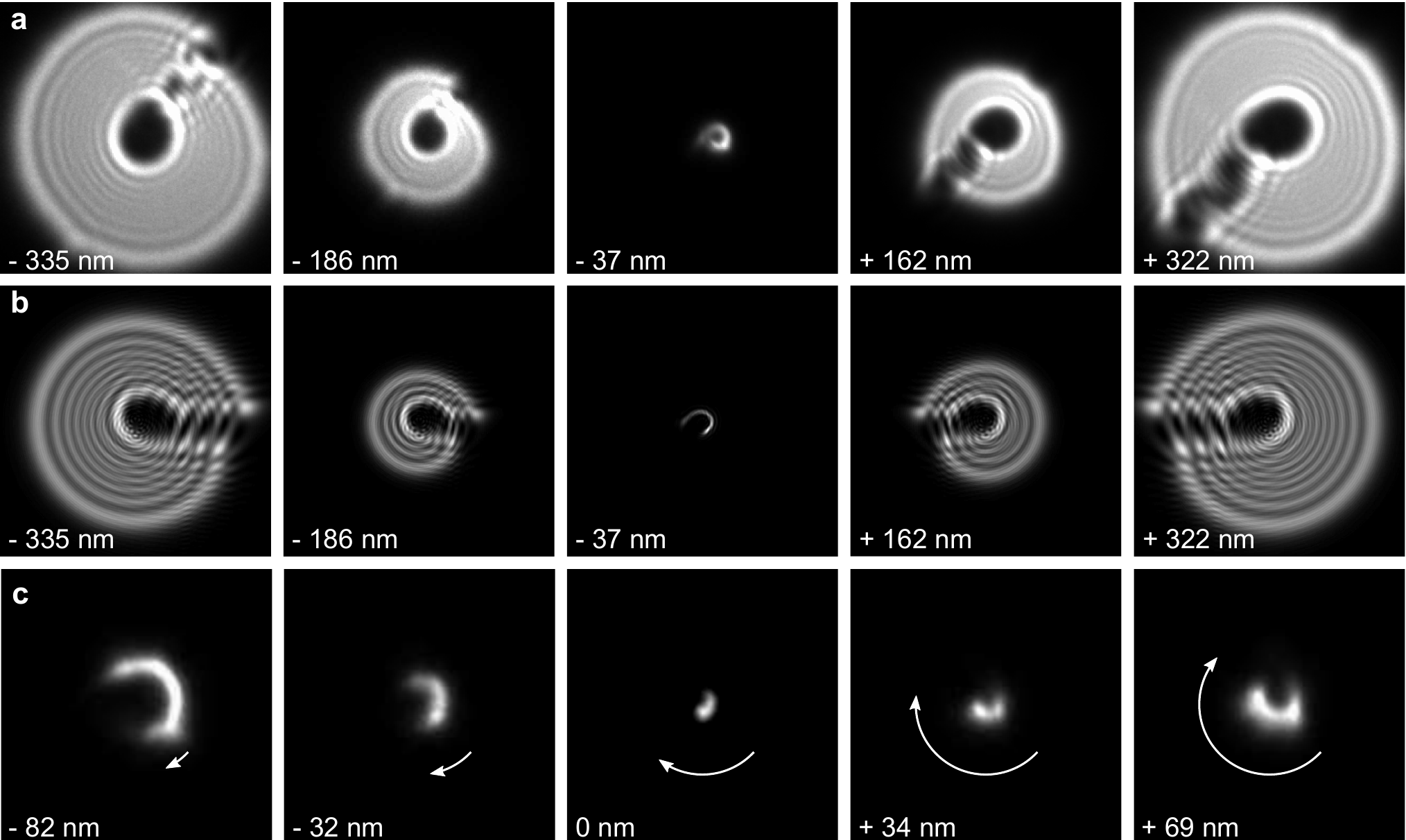}
\caption{\protect\textit{Electron vortex states observed after interaction with the monopole
field:  (a) Through focus series of the needle aperture in the diffraction plane. Note the
persistent dark region in the centre caused by destructive interference typical for vortex waves.
The near-focus central image shows a doughnut like intensity profile, typical of a vortex beam
which opens on one side and indicating a non-integer total orbital angular momentum. (b) Wave
optical simulation obtained by Fourier transforming the complex wave from Fig.2c with a Fresnel
defocus applied. Note the detailed agreement with the experimental figures in (a). A similar
simulation assuming no azimuthal phase is given in Supp. info showing a very different behavior
ruling out the possibility that the black region is caused by shadowing effects from the needle.
Intensity profiles are also given in Supp. info. (c) Through focus series of the beam half cut by
a sharp edge. The rotation of the image along the series proves the presence of a net negative
orbital angular momentum.}}\label{Figure_3}
\end{center}
\end{figure*}

These experiments show that our approximation to a Dirac string indeed provides a magnetic
monopole field. The difference between a true monopole and this approximation lies in the effects
of the flux returning to the needle making the field divergence free again as can be seen from the
defocused images showing Fresnel fringes from the edge and a reconnection of the phase over the
needle (Fig. 3a). This effect is the reason why no forked fringes are observed in the experimental
holograms (Supp. Fig. 3a,b). Detailed holographic simulations showing this subtle reconnection
difference between a true monopole and a Dirac string are shown in Supp. Fig. 7 together with a
sketch explaining the creation of the phase singularities in both cases (Supp. Fig. 8).

The further we go into the far field, the more this effect of the needle disappears and the more
the resulting wave becomes a true electron vortex as if the interaction took place with a real
monopole. It is expected that a needle presenting an integer charge will allow a vortex with
sufficient purity to heal itself, removing this distortion \cite{Bouchal}.

The above experiment shows how quantum experiments with magnetic monopoles are feasible and
provides a very promising way to make electron vortices for applications in electron microscopy
with an almost eight fold gain in beam intensity while avoiding other unwanted beams as compared
to currently used holographic reconstruction methods \cite{Kruit,verbeeck_AtomicScale}. The
current device is static and its magnetic polarization depends entirely on the shape and material
of the needle. However, there are no fundamental obstacles to create a nanoscale solenoid in order
to provide any flux in the Dirac string depending on applied current. This extension would provide
a dynamically switchable source of vortex electrons which would be highly desirable to improve the
speed, flexibility and signal to noise ratio in vortex electron experiments.

Even though an electron microscope was used to conveniently demonstrate the effect, the
consequences of this experiment reach much further as the AB effect is independent on the speed of
the electrons and vortex states can be generated in e.g. solid state systems where conduction
electrons encounter similar monopole fields. Indeed if a sufficiently coherent electron wave
packet in a material would encounter a similar approximate monopole field (e.g. generated by a
ferromagnetic inclusion), it would gain a topologically protected azimuthal phase, possibly
changing its propagation dynamics.

\section{Methods}
The needle was prepared from bulk nickel by a focused ion beam instrument (FIB) using a FEI Helios
Nanolab with Ga ions accelerated to 30 kV. After extraction from the bulk sample, a large nickel
chunk ($\sim$ 10x3x30 $\mu$m$^3$) was welded onto an already prepared conical shaped copper base
and thinned concentrically. The resulting needle is 21.4 $\mu$m long, 700 nm wide at the bottom
and 200 nm at the top. The nickel needle and part of its copper base were then extracted and
sealed over a gold plated SiN film. The needle was precisely placed in order for half of its
length to hang over a 20 $\mu$m circular aperture previously drilled in the Au/SiN film. Electron
holography was performed in Lorentz (field free) mode at 300 kV on the QuAntEM microscope (a
double corrected Titan$^3$ 80-300) and is sketched in Supp. Fig. 1. The M\"{o}llensted biprism
voltage was set to +180 V in order to have a large field of view and good sampling of the
interference fringes. The phase maps were reconstructed from holograms using the standard Fourier
filtering method with a final unwrap step. Such phase maps suffer from the presence of the
electrostatic potential which is corrected by flipping the magnetisation of the needle and
subtracting the phase maps obtained from two opposite magnetisations. The magnetisation state of
the needle is changed by tilting the needle 30 degrees and applying a small magnetic field ($\sim$
0.15 T) by raising the current in the objective lens to 5\% of its full strength. Through focus
series were recorded in the diffraction plane of the Lorentz lens using the highest camera length
available (18 m) and recorded on a CCD mounted at the end of a Gatan Quantum Image Filter. The
Gouy phase experiment was realized in the exact same conditions, cutting the beam with the sharp
edge of an objective aperture to clearly see the rotation effects (Supp. Fig. 1).


\section{Contributions}
A.B. and J.V. conceived the experiment; A.B. designed the sample and carried out the TEM
measurements. All authors contributed to theory, data analysis and writing the paper.

\section{Corresponding author}
Correspondence should be addressed to J.V.:  jo.verbeeck@uantwerp.be.

\section{Acknowledgments}
This work was supported by funding from the European Research Council under the 7th Framework
Program (FP7), ERC grant 246791 COUNTATOMS and ERC Starting Grant 278510 VORTEX. The Qu-Ant-EM
microscope was partly funded by the Hercules fund from the Flemish Government. The authors
acknowledge financial support from the European Union under the Seventh Framework Program under a
contract for an Integrated Infrastructure Initiative. Reference No. 312483-ESTEEM2. R.V.B.
acknowledges a PhD fellowship grant from the FWO (Aspirant Fonds Wetenschappelijk Onderzoek
Vlaanderen).

\newpage

\renewcommand{\figurename}{Supplementary Figure}        
\setcounter{figure}{0}                                  

\parbox{1.8\columnwidth}{
\begin{center}
\Large{Supplementary information: Magnetic monopole field exposed by electrons} 
\end{center}

\centering{A. B\'{e}ch\'{e}, R. Van Boxem, G. Van Tendeloo, J. Verbeeck}

\centering{EMAT, University of Antwerp, Groenenborgerlaan 171, 2020 Antwerp, Belgium}
}

\vspace{1cm}

\begin{minipage}{1.8\columnwidth}
\makebox[\columnwidth]{  
\includegraphics[width=\columnwidth]{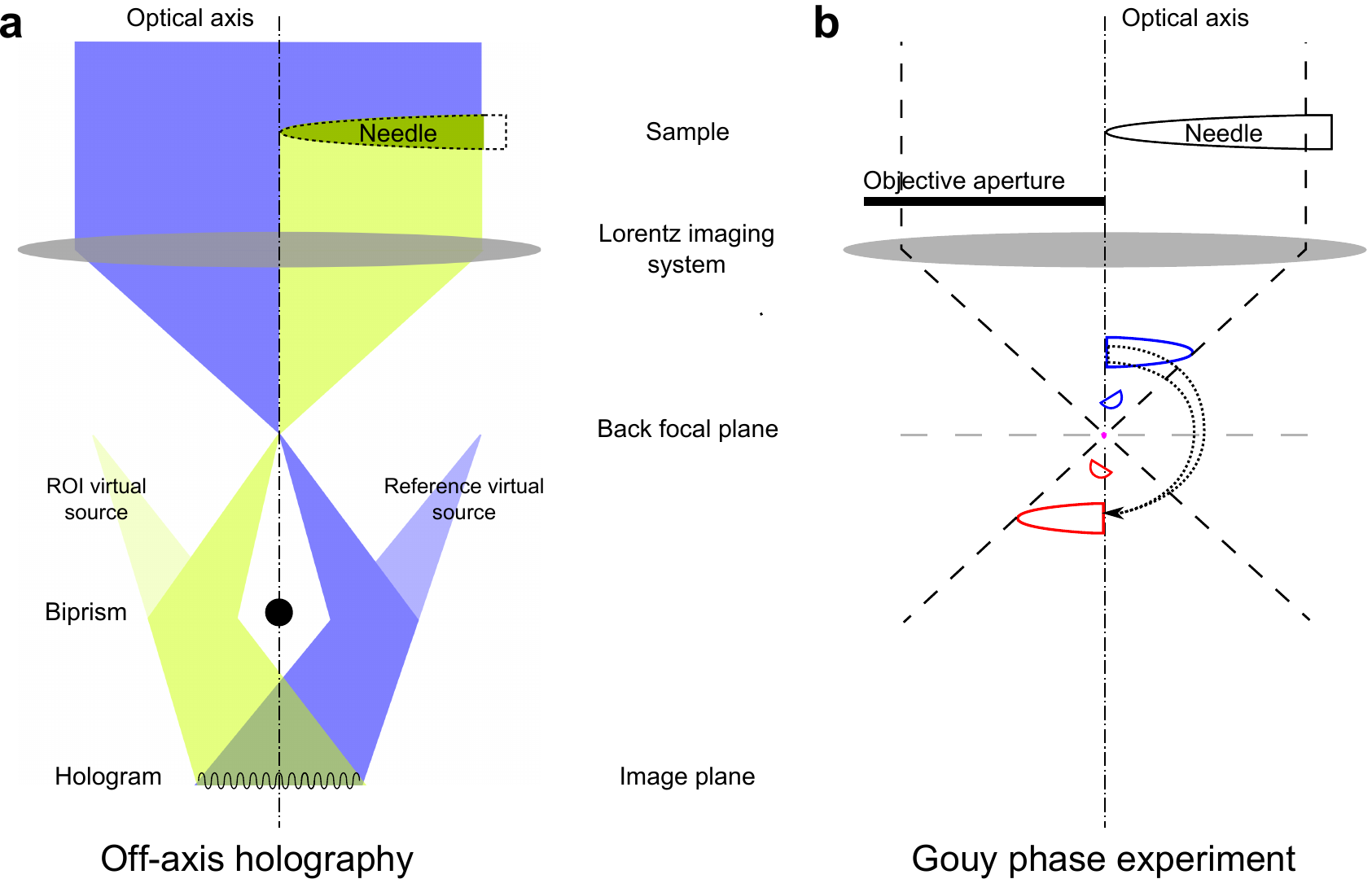}}
\captionof{figure}{\protect\textit{Sketches of experimental setups. (a) Off-axis holography
consists of superimposing a reference wave with a wave passing through the sample. The resulting
fringe pattern, the hologram, carries information about the phase of the wave. (b) The Gouy phase
experiment consists of acquiring a through focus series in the back focal plane. When passing
through the focus point, the shadow image flips. The insertion of a sharp edge blocking part of
the beam, here an objective aperture, allows better visualisation of this image inversion. In the
case of an object illuminated with a vortex beam, as sketched here, the image is not simply
flipped but also rotates when going through focus in a direction given by the sign of the orbital
angular momentum.}}\label{}
\end{minipage}

\begin{figure*}[h]
\begin{center}
\includegraphics[width=1.4\columnwidth]{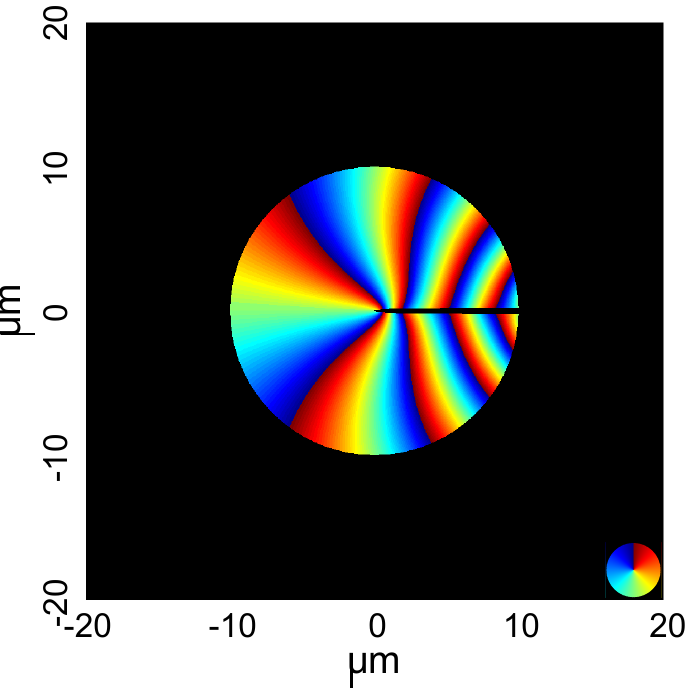}
\caption{\protect\textit{Evolution of the phase around the nickel magnetic needle over the full
aperture. This phase map was calculated from finite element simulations taking into account the
experimental dimensions of the needle (the scale figures a phase change of 2$\pi$). The volume
magnetization value was adapted in the simulation to match the experimental change in phase around
the tip (Fig.2 b). This results in a volume magnetization of 1.72e5 A.m$^{-1}$ instead of 4.88e5
A.m$^{-1}$ expected for bulk Ni. This discrepancy can be attributed to gallium implantation during
the FIB preparation of the needle, reducing the effective volume of Ni which contributes to the
magnetism.}}\label{}
\end{center}
\end{figure*}

\begin{figure*}[h]
\begin{center}
\includegraphics[width=2\columnwidth]{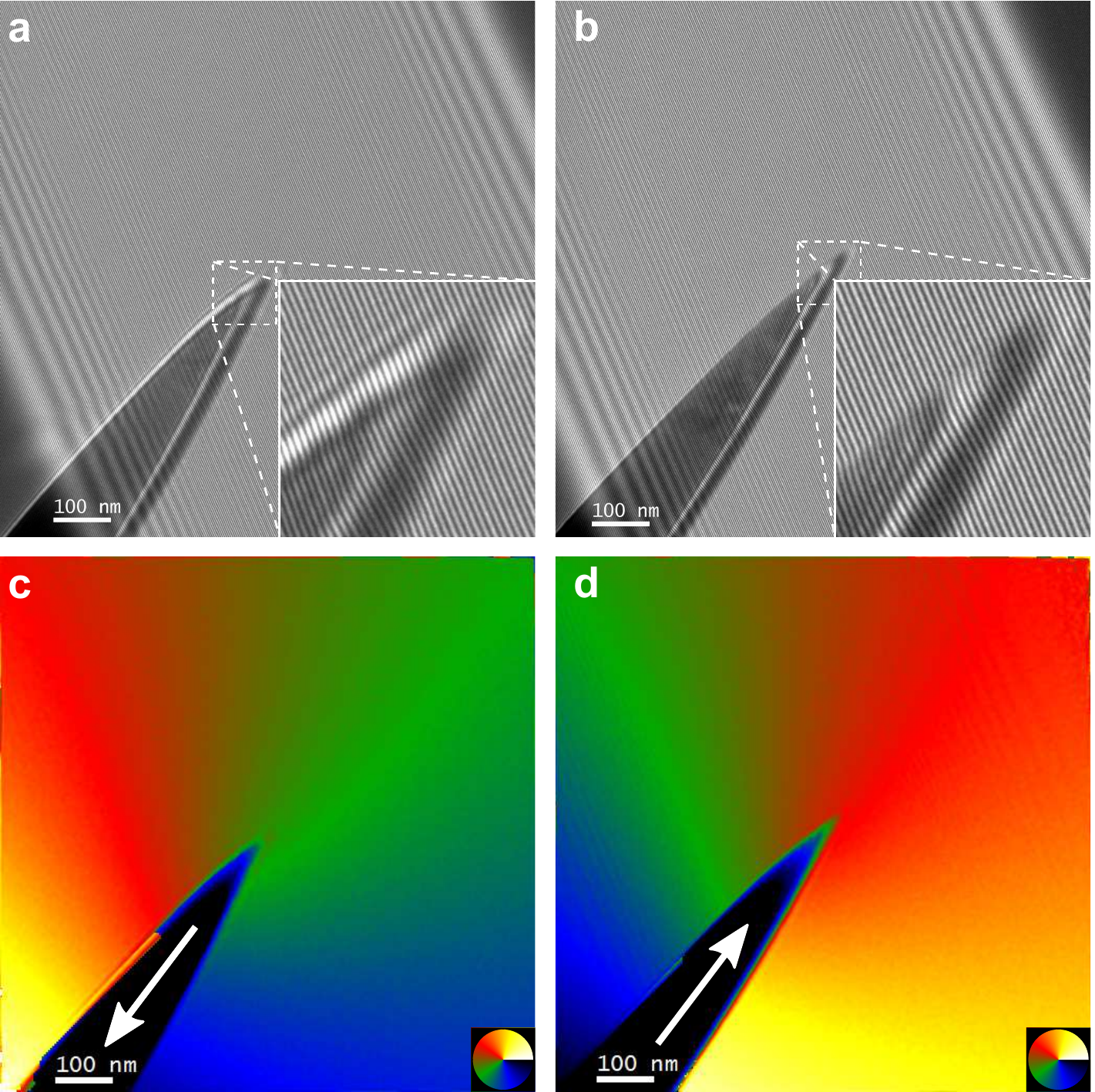}
\caption{\protect\textit{Experimental holograms and phase maps around the needle tip. (a)
Experimental hologram acquired at the tip of the needle in field free conditions (Lorentz mode),
the inset shows an enlarged view of the holographic fringes near the tip of the needle. Note the
absence of forking fringes as further explained in Supplementary Figure 7. (b) Hologram of the
same region taken after changing the direction of the magnetic field together with a magnified
view of the tip as inset, also revealing the absence of forking pattern in the holographic
fringes. To change the direction of the magnetic field, the needle was tilted 30° out of plane and
an out of plane magnetic field of 0.15 T was applied by slightly increasing the strength of the
objective lens. The lens was turned off and the needle was tilted back in plane before acquisition
of the new hologram. (c) Phase map obtained from experimental holograms (a) with the magnetic
field pointing into the tip of the needle. The scale figures a phase change from 0 to 4$\pi$. (d)
Same image taken with the magnetic field pointing out of the needle. The reversal of the
magnetization changes the handedness of the phase distribution as expected. Across the needle, a
black color is used as the phase scale saturates due to the interaction with the mean inner
potential of the material. The color scale is chosen this way to emphasize the magnetically
induced phase. The sum of both figures (c) and (d) was used to remove the influence of the
electrostatic potential and to create Fig 2.b.}}\label{}
\end{center}
\end{figure*}

\begin{figure*}[h]
\begin{center}
\includegraphics[width=2\columnwidth]{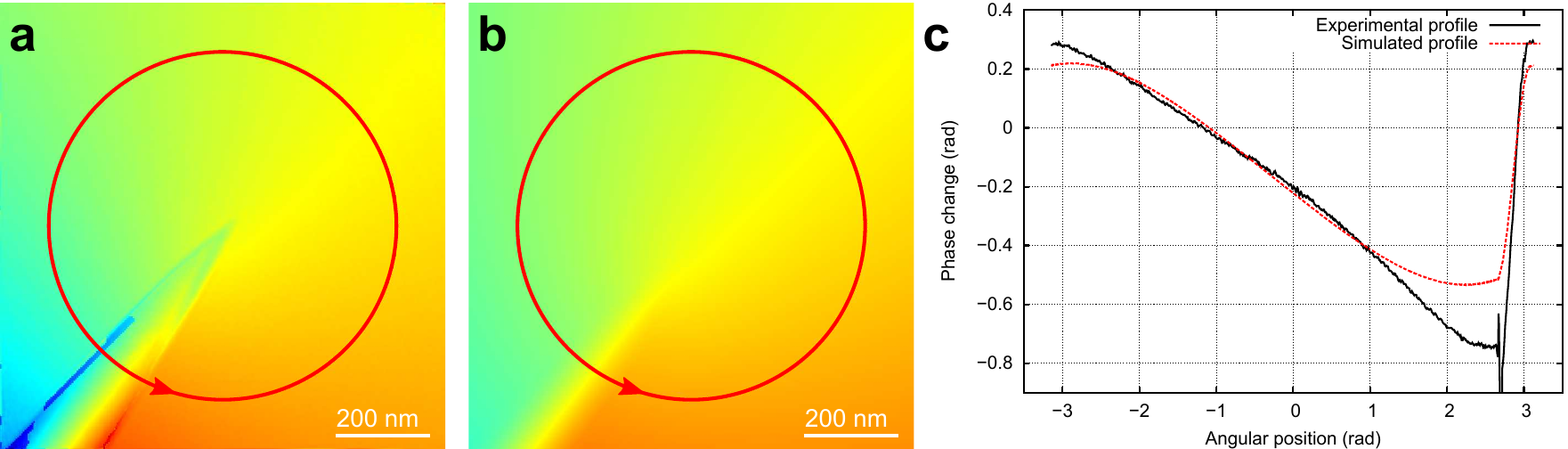}
\caption{\protect\textit{Comparison of experimental and simulated phase shifts. (a) Experimental
phase maps around the tip of the needle. (b) Simulated phase maps corresponding at the same
position as (a). (c) Phase profiles along a circular path around both experimental (a) and
simulated (b) phase maps. The radius of the integrating loop is 195 nm and the profiles starts
from the arrow sketched in (a) and (b). Note the near linear phase decrease in both experimental
and simulated profiles as well as the reconnection of the phase over the needle which is needed
for a divergence free field. This reconnection has a subtle effect on the holograms but the
majority of the electron wave 'sees' an azimuthal phase leading to a clear vortex state even
though the phase is not discontinuous.}}\label{}
\end{center}
\end{figure*}

\begin{figure*}[h]
\begin{center}
\includegraphics[width=1.8\columnwidth]{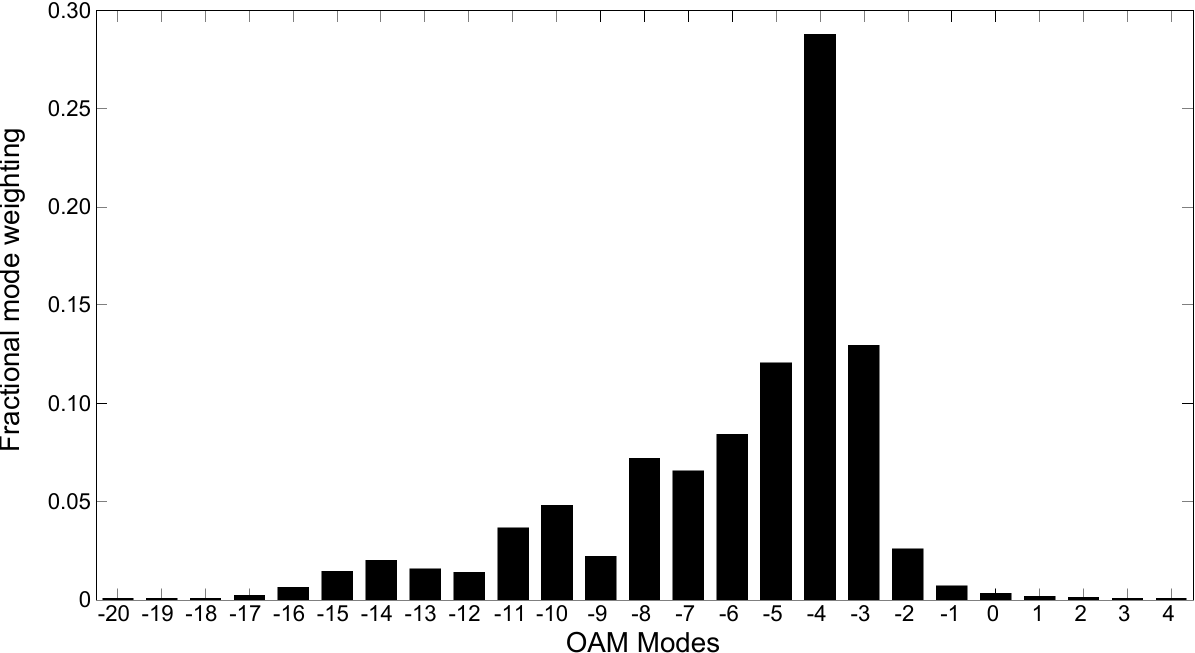}
\caption{\protect\textit{OAM Eigenmodes decomposition of the simulated phase in the full aperture.
The simulated phase presented in Supp. Fig. 2 was used as a source for the decomposition. Note the
clear shift towards negative OAM, indicating the action of the monopole field. The width of the
distribution is caused by the fact that the needle is conical as opposed to an ideal cylindrical
needle which would lead to a single OAM mode.}}\label{}
\end{center}
\end{figure*}

\begin{figure*}[h]
\begin{center}
\includegraphics[width=2\columnwidth]{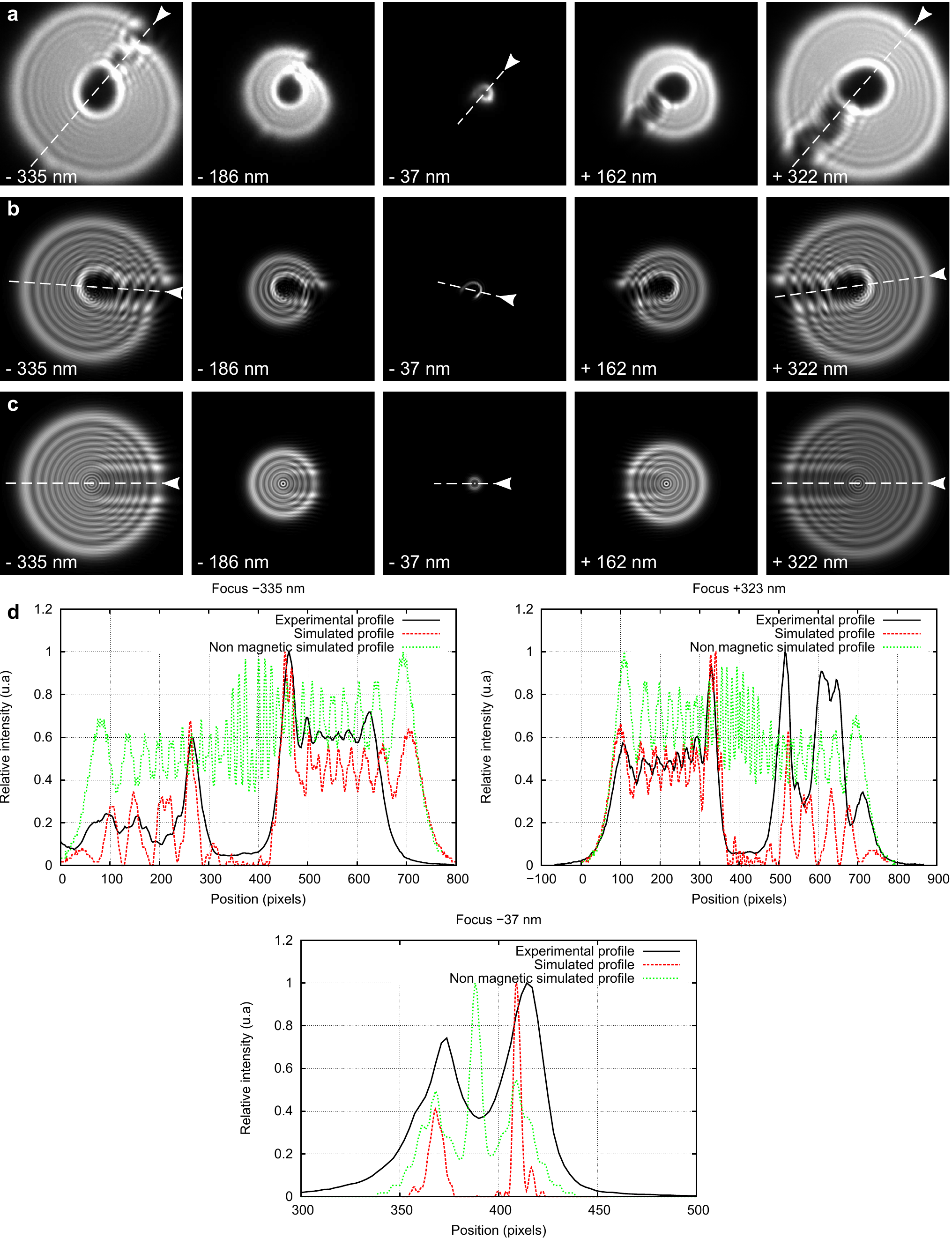}
\caption{\protect\textit{Experimental and simulated through focus series. (a) Experimental through
focus series of the electron intensity near the diffraction plane. (b) Wave optical simulations
assuming the theoretical phase profile from Fig.2c. The simulation is simply a Fourier transform
of the complex wave in the needle plane with a Fresnel defocus term to simulate the effect of
defocus. Note the strong similarity with the experiment. In order to rule out that the black
central part in (a) is caused by a shadowing effect, an alternative simulation assuming a flat
phase (no magnetic effects) profile with a totally blocking needle is given in (c). Indeed a
shadow of the needle is faintly visible but contrary to the experiment the shadow is not truly
black and disappears completely when in focus. These simulations provide strong evidence that a
true vortex state is produced. (d) Intensity profiles are given to further show the essential
effect of the destructive interference near the center of the electron vortex.}}\label{}
\end{center}
\end{figure*}

\begin{figure*}[h]
\begin{center}
\includegraphics[width=2\columnwidth]{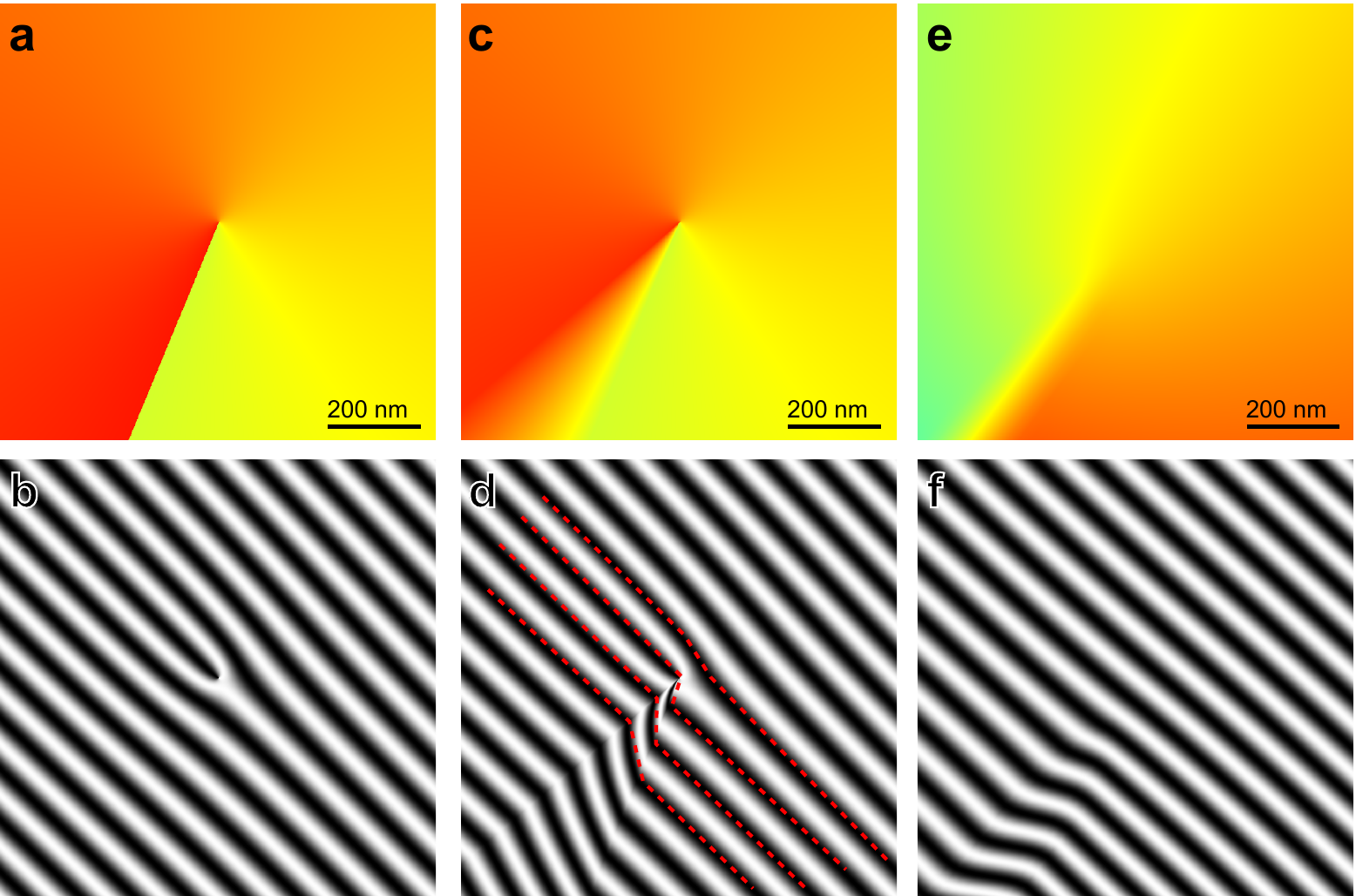}
\caption{\protect\textit{Simulation of the effect of a phase discontinuity on the holographic
fringes. (a) Phase map assuming a perfect azimuthal phase profile with m=1. (b) This phase profile
leads to a typical forked holographic pattern. (c) Phase map presenting a continuous approximation
to phase displayed in (a) with a steep but finite reconnection near the bottom left corner. (d)
Resulting hologram which still resembles (b) but upon detailed inspection the fork has disappeared
and all fringes are single continuous lines as emphasized by the dotted lines. (e) Theoretical
phase shown also in Fig. 2c. (f) Resulting hologram. Note the similarities to (d) without
bifurcations of the fringe lines in agreement with the experimental holograms (Supp. Fig. 3 a,b
insets). These simulations only take into account the magnetic field effect while the
electrostatic potential inside the needle is not simulated for clarity. }}\label{}
\end{center}
\end{figure*}

\begin{figure*}[h]
\begin{center}
\includegraphics[width=1.2\columnwidth]{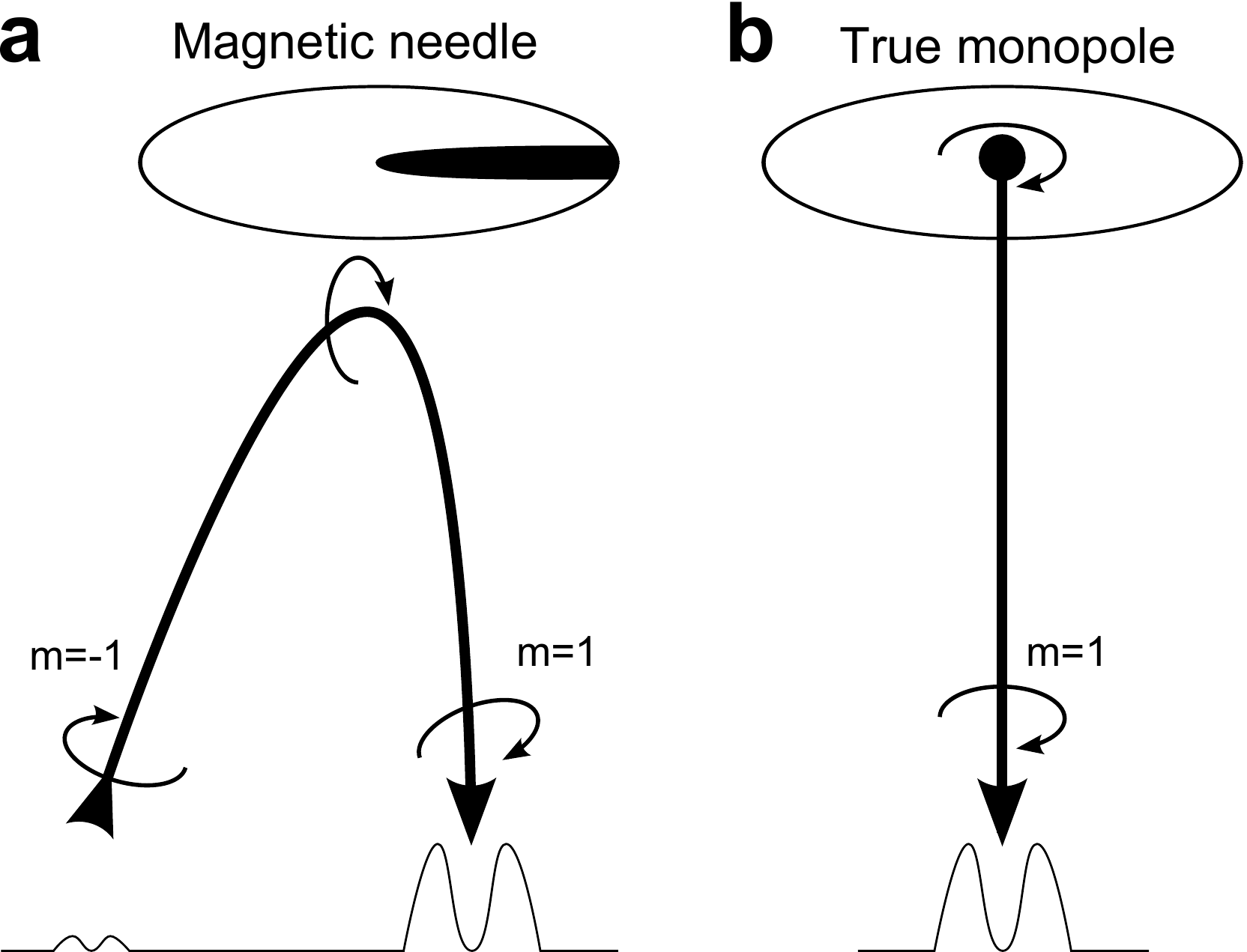}
\caption{\protect\textit{Sketch of the fundamental difference between the interaction with an
approximate monopole field and a true hypothetical magnetic monopole. (a) The approximate monopole
field will imprint a continuously varying phase on the incoming electron wave with an azimuthal
phase profile that is reconnected over the width of the needle. In the immediate vicinity of the
needle there are therefore no phase singularities and no forked holographic fringes would occur if
the phase is measured in this plane (this is a consequence of the field being divergence free).
Nevertheless, propagating the electron wave behind the needle (going from top to bottom), the wave
will develop a vortex-antivortex pair indicated by the black bent arrow. This means that two
singularities arise with opposite winding number. The positive singularity remains on the optical
axis and is surrounded by the majority of electron density in the wave, while the opposite
singularity will split off and shift to high angles (depending on the steepness of the phase
reconnection on the needle) and to a region where the electron density is nearly zero. The fact
that the majority of the electron density is occurring around the main singularity also means that
a real vortex state with an OAM of $\hbar$ is created. The unwanted opposite singularity can be
excluded by making use of an aperture if needed. (b) In case of a true monopole field, the
situation is different and a single phase singularity is created starting from the monopole. This
leads to the formation of an electron vortex wave with OAM of $\hbar$. Both setups therefore have
nearly the same effect on the electron with the exception of the total winding number in the
observation plane which is zero for an approximate monopole field and equals one for the true
monopole field. Putting an aperture to cut out the region of the opposite singularity brings both
cases even closer together. For all practical matters, a monopole field therefore behaves nearly
identical to the field produced by a true monopole as long as we don't observe the near field area
of the scattering.}}\label{}
\end{center}
\end{figure*}

\end{document}